\documentclass[draftclsnofoot,onecolumn]{elsarticle}
\usepackage{amssymb}
\usepackage{amsmath}
\usepackage{setspace}

\usepackage{epsfig}
\usepackage{epstopdf}
\usepackage{color}

\journal{Signal Processing}
\begin{document}

\begin{frontmatter}
\title{Set-Membership Adaptive Constant Modulus Algorithm with a Generalized Sidelobe Canceler and Dynamic Bounds for Beamforming \tnoteref{thanks}}

\author[zju]{Yunlong Cai\corref{cor1}}
\ead{ylcai@zju.edu.cn}
\author[york]{Rodrigo C. de Lamare}
\ead{rcdl500@ohm.york.ac.uk}
\author[zju]{Minjian Zhao}
\ead{mjzhao@zju.edu.cn}

\cortext[cor1]{Corresponding author}
\address[zju]{Department of Information Science and Electronic Engineering, Zhejiang University, Hangzhou 310027, China }
\address[york]{Communications Research Group, Department of Electronics, University of York, YO10 5DD York, U.K.}
\tnotetext[thanks]{This work was supported by the Fundamental Research Funds for the Central Universities, the National Science Foundation of China (NSFC) under Grant $61101103$ and the Scientific Research Fund of Zhejiang Provincial Education Department under Grant Y$201122655$.}

\begin{abstract}
In this work, we propose an adaptive set-membership constant modulus (SM-CM) algorithm with a generalized sidelobe canceler (GSC) structure for blind beamforming.
We develop a stochastic gradient (SG) type algorithm based on the concept of SM filtering for  adaptive implementation.
The filter weights are updated only if the  constraint cannot be satisfied. In addition, we also propose an extension of two schemes of time-varying bounds for beamforming with a GSC structure and incorporate parameter and interference dependence to characterize  the environment which improves the tracking performance of the proposed algorithm in dynamic scenarios.
A convergence analysis of the proposed adaptive SM filtering
techniques is carried out.
Simulation results  show that the proposed adaptive SM-CM-GSC algorithm with dynamic bounds achieves superior performance to previously reported methods at a
reduced update rate.
\end{abstract}


\end{frontmatter}
\section{Introduction}
 Blind  beamforming  has  been widely applied to  system identification, localization and interference suppression   in communications and array processing systems \cite{haykin}-\cite{rude}.
It is often employed with receivers equipped with an antenna array to steer  a directional beampattern towards the desired user and suppress interference without the need for training sequence or pilots in spatial filtering.
In these situations, beam-width and sidelobe levels are the important characteristics of the response and give rise to various beamformer structures, i.e., multiple sidelobe canceller (MSC) and the generalized sidelobe canceller (GSC) \cite{griffiths}-\cite{manolakis}. In particular, the beamformer with a GSC structure that employs a main branch along with a group of auxiliary branches\footnote{In this respect, the interference is assumed to be presented in both main and auxiliary branches, while the desired signal is available in the main branch due to its high gain in the  direction of interest \cite{haykin}.  The auxiliary branches are used to form an estimate of the main branch interference, which is subtracted from the output of the main branch in order to generate the final estimate of the desired signal.}has attracted significant attention.

An important issue is the choice of a suitable criterion for the design of the beamformer.
The constrained minimum variance (MV) and the constrained constant modulus (CM) criteria are considered as the most promising design approaches due to their simplicity and effectiveness.
The  MV-based algorithms are designed in such a way that they attempt to minimize the filter output power while maintaining
a constant response in the direction of a signal of interest  \cite{rcdl2}-\cite{rcdl1}, \cite{honig}, \cite{champagne2}. In fact, the   CM-based algorithms
are based on a criterion that penalizes deviations of the modulus
of the received signal away from a fixed value and forced
to satisfy one or a set of linear constraints such that signals from
the desired user are detected \cite{rude}, \cite{rcdl3}-\cite{cxu21}. The literature indicates that the  CM-based  algorithms outperform the MV-based  algorithms and lead to a solution comparable to that obtained from the minimization of the mean squared error (MSE). Furthermore, the CM-type algorithms are robust against estimation errors and prevent a severe performance degradation in the presence of uncertainties \cite{rcdl3}-\cite{cxu21}. Moreover, it is worth pointing out that the
%
%
beamforming algorithms that incorporate the GSC structures using the CM and MV criteria were proposed in \cite{rude} and \cite{xu}, respectively.
The results showed that the GSC-based blind beamforming algorithms lead to an improved performance compared to the algorithms with a direct-form processor (DFP).

In practice, the beamformer weights must be continually adapted over time in order to cope with changes in the radio signal environment \cite{honig}, \cite{sanderson}. Therefore,
it is preferable to implement blind beamformers with adaptive filtering algorithms such as
stochastic gradient (SG) algorithms  \cite{haykin}.
For this reason the
improvement of blind adaptive SG techniques is an important research
and development topic.
One problem for the adaptive SG algorithms is that
their performance
is strongly dependent on the choice of the step-size value  \cite{haykin}.
Another problem  is the computational complexity associated with the adaptation for every time instant.
Set-membership (SM) filtering  techniques have been proposed to address these  issues \cite{nagaraj}-\cite{sm-ccm}. They specify a bound on the magnitude of the estimation error or the array output, and can reduce the complexity due to data-selective updates.
From \cite{nagaraj}-\cite{sm-ccm}, we can see that the SM filtering techniques are able to achieve a reduction in computation without performance degradation compared to conventional algorithms due to the use of an adaptive step-size for each update.
In particular, the work in \cite{gollamudi2} appears to be the
first approach to combine the SM filtering algorithm with the  CM criterion.
Furthermore, in nonstationary wireless environments, interferers frequently enter and
exit the system, making it very difficult for the SM filtering algorithms to compute
a predetermined error bound and the risk of overbounding and underbounding is significantly increased. Hence, the performance of SM filtering algorithms strongly depends on the error bound specification, which motivates several SM algorithms with time-varying bound schemes \cite{delamare}-\cite{lin1}.

In this work, we present extensions of the methods reported in \cite{delamare} to the GSC structure using blind adaptive set-membership constrained constant modulus (SM-CM)  algorithms for beamforming.
Simulation results  show that the proposed adaptive SM-CM  beamforming algorithm realized in the GSC structure (SM-CM-GSC)
with dynamic bounds achieves superior performance to previously reported methods at a
reduced update rate.
Compared to the existing SM algorithms the contributions of this work are summarized as follows:
 \begin{enumerate}
\item
To the best of our knowledge, there is a very small number of adaptive blind beamforming algorithms  with SM techniques. We develop  an SG-type adaptive  CM beamforming algorithm based on the concept of SM filtering that exploits the GSC structure.
\item The filter weights of the proposed algorithm are updated only if the  constraint cannot be satisfied. Therefore,
 it significantly reduces the computational complexity due to the sparse updates compared to the conventional adaptive  CM-GSC beamforming algorithms.
\item The bounding schemes of the existing SM filtering algorithms cannot be applied to the proposed adaptive blind beamforming algorithm in nonstationary scenarios. We propose   two schemes of time-varying bounds for beamforming with a GSC structure and incorporate parameter and interference dependence to characterize  the environment for improving the tracking performance  in dynamic scenarios.
\item
A convergence analysis of the proposed adaptive SM filtering
techniques is carried out and analytical expressions to predict
the steady-state MSE are obtained.
\end{enumerate}

In this paper, the superscripts $(.)^{T}$, $(.)^{*}$, $(.)^{-1}$, and $(.)^{H}$
denote transpose, element-wise conjugate, matrix inverse,
 and Hermitian transpose, respectively.
Bold symbols denote matrices or vectors.
The symbols $E[.]$, $|.|$,
$||.||$,
$\mathbf{I}$ and $\mathbf{0}$ represent the  expectation operator, the norm
of a scalar, the norm of a vector,   an identity
matrix of appropriate dimension and a zero vector of appropriate dimension, respectively.

The remainder of this paper is organized as follows: we briefly describe
a system model for beamforming and
 the design of CM beamformers with a GSC structure in Section \ref{Section2:system}.
The SM filtering framework and the adaptive blind SM-CM-GSC algorithm are introduced in Section \ref{Section3:smadapive}.
Section \ref{Section4:timevaryingbounds} introduces two strategies to compute time-varying bounds for the proposed algorithms.
 Convergence  analysis of the resulting
algorithm and the analytical formulas to predict the
steady-state MSE are developed in Section \ref{Section5:analysis}. The simulation results are presented in Section \ref{Section6:simulations}. Finally,
Section \ref{Section7:Conclusions} draws the conclusions.

\section{System Model and  Linearly CM-GSC Beamformer}
\label{Section2:system}

%
Let us suppose that $q$ narrowband signals impinge on a uniform linear array (ULA) of $m$ ($m\geq q$) sensor elements. The sources are assumed to be in the far field with direction of arrivals (DOAs)
$\theta_{0},\ldots, \theta_{q-1}$. The $i$th snapshot's received vector $\mathbf{r} \in \mathcal{C}^{m\times 1}$ can be modeled as
\begin{equation}
\mathbf{r}(i)=\mathbf{A}(\mathbf{\mbox{\boldmath$\theta$}})\mathbf{b}(i)+\mathbf{n}(i),
\end{equation}
where $\mbox{\boldmath$\theta$}=[\theta_{0}, \ldots, \theta_{q-1}]^{T}\in \mathcal{R}^{q\times 1}$ is the vector with the DOAs of the signals, $\mathbf{A}(\mbox{\boldmath$\theta$})=[\mathbf{a}(\theta_{0}), \ldots, \mathbf{a}(\theta_{q-1})]\in \mathcal{C}^{m\times q}$ comprises the normalized signal steering vectors $\mathbf{a}(\theta_{k})\in \mathcal{C}^{m\times 1}$
\begin{equation}
\mathbf{a}(\theta_{k})=\frac{1}{\sqrt{m}}[1, e^{-2\pi j \frac{u}{\lambda_{c}}\cos (\theta_{k})}, \ldots,e^{-2\pi j (m-1)\frac{u}{\lambda_{c}}\cos (\theta_{k})} ]^{T},
\end{equation}
%
 where $k=0,\ldots, q-1$, $\lambda_{c}$ is the wavelength, $u$ ($u=\frac{\lambda_{c}}{2}$ in general) is the inter-element distance of the ULA. To avoid mathematical ambiguities, the
steering vectors $\mathbf{a}(\theta_{k})$ are assumed to be linearly independent,
 $\mathbf{b}(i)=[b_{0}(i),b_{1}(i),\ldots,b_{q-1}(i)]^{T}$
 is the source data vector, where we assume that the signals are independent and identically distributed (i.i.d) random variables with equal probability from the set $\{\pm 1\}$. The vector $\mathbf{n} \in \mathcal{C}^{m\times 1}$ is  a  Gaussian noise with  $E[\mathbf{n}\mathbf{n}^{H}]=\sigma^{2}_{n}\mathbf{I}$, where $\sigma^{2}_{n}$ denotes the noise variance.
 In this work, we assume that $\theta_{0}$ corresponds to the direction of the desired user with respect to the antenna arrays and is known beforehand by the beamformer. In practice, $\theta_{0}$ can be estimated by DOA estimation algorithms.
The output of a narrowband GSC beamformer  is given by
\begin{equation}
y(i)=\mathbf{\tilde{w}}^{H}(i)\mathbf{r}(i),
\end{equation}
where $\mathbf{\tilde{w}}(i)=v\mathbf{a}(\theta_{0})- \mathbf{B}^{H}\mathbf{w}(i)$,
$\mathbf{B}\in\mathcal{C}^{(m-1)\times m}$ denotes the signal blocking matrix\footnote{
It is obtained by the singular value decomposition (SVD) or the QR decomposition algorithms and collecting eigenvectors corresponding to null eigenvalues \cite{goldstein}.
Thus, $\mathbf{B}\mathbf{a}(\theta_{0})=\mathbf{0}$ means that the term $\mathbf{B}$ effectively blocks any signal coming from the look direction $\theta_{0}$.},
$\mathbf{w}(i)=[w_{1}, \ldots, w_{m-1}]^{T}\in \mathcal{C}^{(m-1) \times 1}$ is the complex weight vector of the filter and $v$ is a real constant. That is to say, the GSC structure consists of a main branch and an auxiliary branch. The output of the main branch is $v\mathbf{a}^{H}(\theta_{0})\mathbf{r}(i)$, and the output of the auxiliary branch is $\big(\mathbf{B}^{H}\mathbf{w}(i)\big)^{H}\mathbf{r}(i)$.
The  auxiliary branch is employed to form an estimate of the main branch interference that can be used for cancelation.

The  CM-GSC optimization problem determines the filter parameters $\mathbf{w}(i)$ by solving
\begin{equation}
\textrm {minimize} \quad J_{CM}(\mathbf{w}(i))=E\Big[ \big( | \mathbf{\tilde{w}}^{H}(i)\mathbf{r}(i) |^2-1  \big)^{2} \Big].\label{eq:cm1}
\end{equation}
The objective of (\ref{eq:cm1}) is to minimize the expected deviation of the squared modulus of the beamformer output to a constant
while maintaining the contribution from $\theta_{0}$ constant, i.e. $\mathbf{\tilde{w}}^{H}(i)\mathbf{a}(\theta_{0})=(v\mathbf{a}(\theta_{0})- \mathbf{B}^{H}\mathbf{w}(i) )^{H}\mathbf{a}(\theta_{0})=v$. The  CM-GSC design can have its convexity enforced by adjusting the parameter $v$, note that the detailed analysis of the optimization problem is shown in the Appendix. The  CM-GSC filter expression that iteratively solves the problem in (\ref{eq:cm1}) is given by
\begin{equation}
\mathbf{w}(i+1)=\big(E[|y(i)|^2\mathbf{B}\mathbf{r}(i)\mathbf{r}^{H}(i)\mathbf{B}^{H}]\big)^{-1}E[(v\mathbf{a}^{H}(\theta_{0})y^{*}(i)\mathbf{r}(i)-1)^{*}y^{*}(i)\mathbf{B}\mathbf{r}(i)],\label{eq:itercmgsc1}
\end{equation}
where $y(i)=\mathbf{\tilde{w}}^{H}(i)\mathbf{r}(i)=v\mathbf{a}^{H}(\theta_{0})\mathbf{r}(i)- \mathbf{w}^{H}(i)\mathbf{B}\mathbf{r}(i)$.
It should be remarked that the expression in (\ref{eq:itercmgsc1}) is a function of previous values of  filter $\mathbf{w}(i)$ and therefore must be iterated in order to reach a solution.
However, the method of computing (\ref{eq:itercmgsc1}) is not practical in wireless communications applications with mobile users and nonstationary interferers, and hence an adaptive implementation is needed.
The SG algorithm is one of the most widely used adaptive algorithms,
but one problem with this algorithm is the computational complexity
related to the adaptation for each snapshot.
In order to reduce the update rate and improve the convergence performance,
we will introduce the proposed  adaptive low-complexity SM beamforming algorithm in the following section.


\section{Proposed Adaptive SM Technique}
\label{Section3:smadapive}

In this work, we develop the SM-CM-GSC adaptive SG algorithm that updates the filter weights only if the bound constraint $e^2(i)\leq \gamma^2$ cannot be satisfied, where $e(i)=|\mathbf{\tilde{w}}^{H}(i)\mathbf{r}(i)|^2-1$ denotes the prediction error and $\gamma$ denotes a specified bound.
The solution of the proposed algorithm is a set in the parameter space \cite{rcdl14}, which includes
some estimates that satisfy the bound constraint  corresponding to different $\mathbf{r}$ for different time instants.
%
\begin{figure}[p]
\centering \scalebox{0.9}{\includegraphics{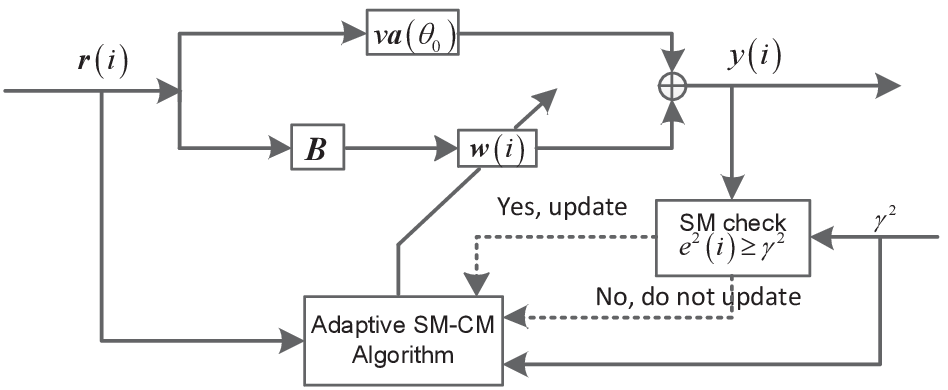}}
\caption{Proposed SM-CM-GSC adaptive beamformer structure.}
\label{fig:smcmgscbd}
\end{figure}

\subsection{Proposed SM Framework}

Let us define a sample space $\mathbb{S}$ that contains all possible data  $\{ \mathbf{r}\}$.
Then, we define the \textit{feasibility set} $\mathcal{Q}$ as
\begin{equation}
\mathcal{Q}=\bigcap_{\mathbf{r}\in \mathbb{S}} \{\mathbf{w} \in \mathcal{C}^{(m-1)\times 1}: (|\mathbf{\tilde{w}}^{H}\mathbf{r}|^2-1)^2\leq \gamma^{2}\},
\end{equation}
which contains the values that fulfill the error bound.

We apply the feasibility set to a time-varying scenario; therefore, it contains all estimates that fulfill the  bound constraint at the $i$th time instant. This set is termed the \textit{constraint set} and is given by
\begin{equation}
\mathcal{H}_{i}=\{\mathbf{w} \in \mathcal{C}^{(m-1)\times 1}: (|\mathbf{\tilde{w}}^{H}\mathbf{r}(i)|^2-1)^2\leq \gamma^{2}  \}.
\end{equation}
Our aim is to develop an adaptive algorithm that updates the parameters such that it will always remain within the constraint set.
As depicted in Fig. \ref{fig:smcmgscbd}, the proposed adaptive scheme introduces the principle of the SM filtering technique into the blind CM beamforming algorithm with a GSC structure. Thus,
it operates with respect to certain snapshots and therefore  has a reduced computational complexity.  Furthermore, the data-selective updates will lead to highly effective variable step-size for the SG-based SM beamforming algorithm. In the following, we will describe the proposed blind adaptive algorithm in detail.

\subsection{Proposed SM-CM-GSC Adaptive Algorithm}

We devise a gradient descent strategy to compute the filter weight vector $\mathbf{w}$
that minimizes the instantaneous  CM-GSC cost function, the adaptation is required when the square of the error $e^{2}(i)$ exceeds a specified error bound $\gamma^{2}(i)$. Note that the bound here can be assumed to be time-varying and based on the estimated parameters of the filter weight vector, and the time-varying bound schemes will be addressed in the next section.
The problem is formulated as follows,
\begin{equation}
\textrm {minimize} \quad \quad J_{CM}=  \big( | \mathbf{\tilde{w}}^{H}(i)\mathbf{r}(i) |^2-1  \big)^{2},\label{eq:cm2inal}
\end{equation}
\begin{equation}
\textrm {whenever} \quad e^{2}(i)> \gamma^{2}(i).
\end{equation}
We consider the following gradient search procedure:
\begin{equation}
\mathbf{w}(i+1)=\mathbf{w}(i)-\mu(i) \frac{\partial J_{CM}} {\partial \mathbf{w}^{*}},
\end{equation}
where $\mu(i)$ is the effective variable step-size.
By taking the gradient of (\ref{eq:cm2inal}) with respect to $\mathbf{w}^{*}$ we have $\frac{\partial J_{CM}} {\partial \mathbf{w}^{*}}=-|y(i)|^2\mathbf{B}\mathbf{r}(i)y^{*}(i)+\mathbf{B}\mathbf{r}(i)y^{*}(i)$. Then, we obtain the following SG algorithm
\begin{equation}
\mathbf{w}(i+1)=\mathbf{w}(i)-\mu(i)\Big(\mathbf{B}\mathbf{r}(i)y^{*}(i)-|y(i)|^2\mathbf{B}\mathbf{r}(i)y^{*}(i)  \Big).\label{eq:ccm3}
\end{equation}
The variable step-size value will attempt to find the shortest path from $\mathbf{w}(i)$ to the bounding hyperplane of $\mathcal{H}_{i}$ in accordance with the principle of minimal disturbance. In other words, $\mathbf{w}(i+1)$ is the projection of $\mathbf{w}(i)$ on $\mathcal{H}_{i}$.
However, if $\mathbf{w}(i)\in \mathcal{H}_{i}$, we can see that the error bound constraint is satisfied; therefore, no update is necessary, and $\mathbf{w}(i+1)=\mathbf{w}(i)$.
Note that the constraint set comprises of two parallel hyper-strips in the parameter space. Based on the constraint $e^2(i)> \gamma^2(i)$, we consider the following two cases for update:
1) $|y(i)|> \sqrt{1+\gamma(i)}$ and 2) $|y(i)|< \sqrt{1-\gamma(i)}$, and obtain the following expression for $\mu(i)$:
\begin{equation}
\mu(i)=
\left
 \{
\begin{array}{cc} \Big(1-\frac{\sqrt{1+\gamma(i)}}{|y(i)|} \Big) \frac{1}{(\mathbf{r}^{H}(i)\mathbf{B}^{H}|y(i)|^{2}-\mathbf{r}^{H}(i)\mathbf{B}^{H})\mathbf{B}\mathbf{r}(i)} & \textrm {if} \quad | y(i)| > \sqrt{1+\gamma(i)} \\
 \Big(1-\frac{\sqrt{1-\gamma(i)}}{|y(i)|}   \Big) \frac{1}{(\mathbf{r}^{H}(i)\mathbf{B}^{H}|y(i)|^{2}-\mathbf{r}^{H}(i)\mathbf{B}^{H})\mathbf{B}\mathbf{r}(i)}  &    \textrm {if} \quad | y(i)|  < \sqrt{1-\gamma(i)} \\
0 &   \textrm {otherwise}
\end{array}
\right.
\label{eq:smcmgsc2}
\end{equation}
where the derivations are detailed in \ref{section10:appendixc}.
The proposed SM-CM-GSC adaptive algorithm which consists of  equations (\ref{eq:ccm3}) and (\ref{eq:smcmgsc2}) updates the filter vector $\mathbf{w}(i)$ over time in a manner to converge to the optimum filter weight vector corresponding to (\ref{eq:itercmgsc1}).
The SM filtering technique with a time-varying bound is employed to determine a set of estimates $\{\mathbf{w}(i) \}$ that satisfy the bounded constraint.

Note that the blocking matrix  $\mathbf{B}$ has no particular structure if the SVD or QR decomposition is employed. Therefore, the complexity of computing the error $y(i)$ is  high, since computing the error involves carrying out the multiplication $\mathbf{B}\mathbf{r}(i)$.
%
However, there are several ways to easily bypass this computational problem \cite{werner1}-\cite{tsengcy}. One  method   is  the application of the correlation subtractive structure (CSS) \cite{werner1}, \cite{ricks}. The commonly used blocking matrix with the CSS implementation is given by  $\mathbf{B}=\mathbf{I}-\mathbf{a}(\theta_{0})\mathbf{a}^{H}(\theta_{0})$. By using
 the particular structure of the blocking matrix, we can compute the error with a linear complexity.
For each snapshot, the conventional adaptive  CM-GSC beamforming algorithm requires $3m$ multiplications and $3m-1$ additions, while
  the proposed SM-CM-GSC adaptive beamforming algorithm requires $2m+\eta m$ multiplications and $2(m-1)+\eta(m+1)$ additions, where $0<\eta\leq1$ denotes the update rate.
  In particular, for a configuration with $m=40$ and $\eta=20\%$, the number of multiplications for the conventional  CM-GSC and the proposed SM-based algorithms are $120$ and $88$, respectively. The number of additions for them are $119$ and $86$, respectively.
  It is worth  mentioning that the computational complexity is reduced significantly due to the data-selective updates.

\section{Time-Varying Bound Schemes}
\label{Section4:timevaryingbounds}

The bound of  SM filtering algorithms  is an important quantity to measure the quality of the estimates that could be included in the constraint set.
In \cite{Bhotto}, \cite{lin1}, several predetermined bounding schemes have been reported for development of the adaptive SM filters, which achieve reduced complexity without performance degradation.
However, in a nonstationary scenario, they are impractical to reflect the time-varying nature of the environment and may result in poor convergence and tracking performance.
To the best of our knowledge, there is a very small number of works employing time-varying bounds for SM filters.
In this work,
we present extensions of the methods reported in \cite{delamare} to the GSC scheme to compute the time-varying error bound $\gamma(i)$, which is a single coefficient to check if the filter update is carried out or not.

\subsection{Parameter Dependent Bound (PDB)}

The proposed time-varying bound schemes can increase the convergence and tracking performance. The first scheme is called parameter dependent bound (PDB).
It computes a bound for the SM-CM-GSC adaptive algorithm and is given by
\begin{equation}
\gamma(i+1)=(1-\rho)\gamma(i)+\rho\sqrt{\lambda ||\mathbf{\tilde{w}}(i) ||^{2} \hat{\sigma}^{2}_{n}(i) },\label{eq:bound1}
\end{equation}
where $\rho$ is a forgetting factor parameter that should be set to guarantee a proper time-averaged estimate of the evolution of the power of GSC beamforming vector $\mathbf{\tilde{w}}(i)=v\mathbf{a}(\theta_{0})- \mathbf{B}^{H}\mathbf{w}(i)$, $\lambda$ ($\lambda>1$) is a tuning coefficient and $\hat{\sigma}^{2}_{n}(i)$ is an estimate of the noise power.
We assume that the noise power is known beforehand at the receiver.
The time-varying bound provides a smoother
evolution of the weight vector trajectory and thus avoids too high or low values of the squared norm of the weight vector.
It establishes a relation between the estimated parameters and the environmental coefficients.

\subsection{Parameter and Interference Dependent Bound (PIDB)}
\begin{figure}[!hhh]
\centering \scalebox{0.9}{\includegraphics{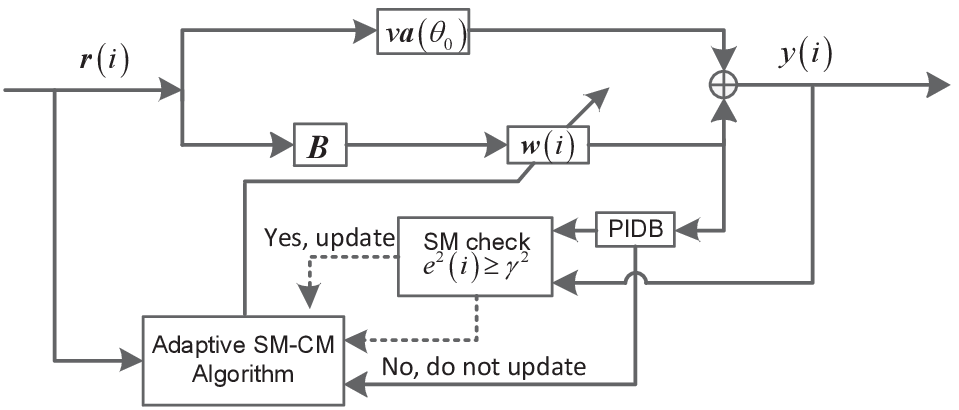}}
\caption{Proposed SM-CM-GSC adaptive beamformer structure with PIDB scheme.}
\label{fig:smcmgscpidbbd}
\end{figure}

The second time-varying bound scheme has a slightly increased complexity compared to the PDB scheme. It combines the PDB with the interference estimation that is provided by the auxiliary branch of the GSC structure.
It provides more information about the environment for parameter estimation  and has an improved performance. We refer to it as parameter and interference dependent bound (PIDB), and it is an extension of \cite{delamare} for beamforming design with the  CM-GSC criterion. The proposed SM-CM-GSC adaptive beamforming algorithm with the PIDB structure is shown in Fig. \ref{fig:smcmgscpidbbd}.

Since the matrix $\mathbf{B}$ blocks the signal which comes from the desired direction, the auxiliary branch of the GSC structure generates the estimate of the interference and the noise. The  power of the interference and the noise is given by
\begin{equation}
E[|\mathbf{w}^{H}(i)\mathbf{B}\mathbf{r}(i)|^2]=\mathbf{w}^{H}(i)\mathbf{B}\bigg(\sum^{q-1}_{k=1}\mathbf{a}(\theta_{k})\mathbf{a}^{H}(\theta_{k})+\sigma^{2}_{n}\mathbf{I} \bigg)\mathbf{B}^{H}\mathbf{w}(i).\label{eq:interferandnoise}
\end{equation}
By using time averages of the instantaneous values, we can obtain an estimate of (\ref{eq:interferandnoise}), which is
\begin{equation}
\nu(i+1)=(1-\rho)\nu(i)+\rho |\mathbf{w}^{H}(i)\mathbf{B}\mathbf{r}(i)|^{2}, \label{eq:bound2}
\end{equation}
the component $\nu(i)$  performs the estimate of the interference and the noise power,  $\rho$ is a forgetting factor to ensure a proper time-averaged estimate.
By incorporating the information of the interference and noise power into the bounding scheme we have the PIDB expression
\begin{equation}
\gamma(i+1)=(1-\rho)\gamma(i)+\rho\big(\sqrt{\psi \nu(i)}+\sqrt{\lambda ||\mathbf{\tilde{w}}(i) ||^{2} \hat{\sigma}^{2}_{n}(i) }\big), \label{eq:bound3}
\end{equation}
where $\psi$ is a weighting parameter that should be set,
 note that update equation (\ref{eq:bound2}) avoids instantaneous values that are undesirably too high or too low, and thus avoids inappropriate estimates of $\gamma(i)$. Compared with (\ref{eq:bound1}), the PIDB involves
the  estimate of the interference and the noise power and provides more information to track the characteristics of the environments.




\section{Analysis of the Proposed Algorithm}
\label{Section5:analysis}

In this section, we investigate the convergence behavior of our proposed SM schemes when used in the adaptive  CM-GSC beamforming algorithm in terms of the steady-state excess MSE.
The nonlinearities in the update equations of the  CM-GSC beamformer  usually lead to significant difficulties in the study of their performance.
We use a very efficient approach named energy conservation principle \cite{sayed1}-\cite{yousef} which overcomes many of these difficulties.

\subsection{The Range of Step-Size Values for Convergence}

In this part, we discuss the range of the step-size values for convergence. In order to do the analysis, we need to write the proposed beamforming filter weights update equation. Let us recall (\ref{eq:ccm3}), by multiplying $-\mathbf{B}^{H}$ and adding $v\mathbf{a}(\theta_{0})$ on both sides  we have
\begin{equation}
\begin{split}
\mathbf{\tilde{w}}(i+1)&=\mathbf{\tilde{w}}(i)-\mu(i)e(i)\mathbf{r}^{H}(i)\mathbf{\tilde{w}}(i)\mathbf{B}^{H}\mathbf{B}\mathbf{r}(i)\\&
=(\mathbf{I}-\mu(i)e(i)\mathbf{d}(i)\mathbf{r}^{H}(i))\mathbf{\tilde{w}}(i).
\end{split}
\end{equation}
Further, we obtain
\begin{equation}
\begin{split}
\mbox{\boldmath$\varepsilon$}(i+1)&=\mathbf{\tilde{w}}_{opt}-\mathbf{\tilde{w}}(i+1)
\\&=(\mathbf{I}-\mu(i)e(i)\mathbf{d}(i)\mathbf{r}^{H}(i))\mbox{\boldmath$\varepsilon$}(i)+
\mu(i)e(i)\mathbf{d}(i)\mathbf{r}^{H}(i)\mathbf{\tilde{w}}_{opt},\label{eq:weighterror}
\end{split}
\end{equation}
where $\mathbf{d}(i)=\mathbf{B}^{H}\mathbf{B}\mathbf{r}(i)$, $\mathbf{\tilde{w}}_{opt}=v\mathbf{a}(\theta_{0})-\mathbf{B}^{H}\mathbf{w}_{opt}$ denotes the optimum beamformer, and $\mathbf{w}_{opt}$ denotes the optimum filter for $\mathbf{w}$.
By taking expectations on both sides of (\ref{eq:weighterror}) we have
\begin{equation}
E[\mbox{\boldmath$\varepsilon$}(i+1)]=(\mathbf{I}-E[\mu(i)]\mathbf{R}_{dr}(i))E[\mbox{\boldmath$\varepsilon$}(i)],\label{eq:weighterror2}
\end{equation}
where $\mathbf{R}_{dr}(i)=E[e(i)\mathbf{d}(i)\mathbf{r}^{H}(i)]$ and $\mathbf{R}_{dr}(i)\mathbf{\tilde{w}}_{opt}\approx\mathbf{0}$ \cite{honig}.
Therefore, it can be concluded that $\mathbf{\tilde{w}}$ converges to $\mathbf{\tilde{w}}_{opt}$ and (\ref{eq:weighterror2}) is stable if and only if
$\Pi^{\infty}_{i=0}(\mathbf{I}-E[\mu(i)]\mathbf{R}_{dr})\rightarrow 0$, which is a necessary and sufficient condition for $\lim_{i\rightarrow \infty}E[\mbox{\boldmath$\varepsilon$}(i)]=\mathbf{0}$ and $E[\mathbf{\tilde{w}}(i)]\rightarrow \mathbf{\tilde{w}}_{opt}$. For stability, a sufficient condition for (\ref{eq:weighterror2}) to hold implies that \cite{haykin}
\begin{equation}
0\leq E[\mu(\infty)] < \min_{k} \frac{2}{|\lambda^{dr}_{k}|},
\end{equation}
 where $\lambda^{dr}_{k}$ is the $k$th eigenvalue of $\mathbf{R}_{dr}$ that is not real since it is not symmetric.

\subsection{Steady-State  Analysis for Excess MSE}

Let us define the MSE at time index $i$ using the following expression
\begin{equation}
\begin{split}
\xi(i)&=E\big[|b_{0}(i)-\mathbf{\tilde{w}}^{H}(i)\mathbf{r}(i)|^{2} \big]\\&
=E\big[|b_{0}(i)-(\mathbf{\tilde{w}}_{opt}-\mbox{\boldmath$\varepsilon$}(i))^{H}\mathbf{r}(i)|^{2} \big]\\&
=\xi_{min}+E[|e_{a}(i)|^2]+\mathbf{a}^{H}(\theta_{0})E[\mbox{\boldmath$\varepsilon$}(i)]+E[\mbox{\boldmath$\varepsilon$}^{H}(i)]\mathbf{a}(\theta_{0})
\\&\quad-E[\mathbf{\tilde{w}}^{H}_{opt}\mathbf{r}(i)\mathbf{r}^{H}(i)\mbox{\boldmath$\varepsilon$}(i)]-E[\mbox{\boldmath$\varepsilon$}^{H}(i)\mathbf{r}(i)\mathbf{r}^{H}(i)\mathbf{\tilde{w}}_{opt}]
\end{split}
\end{equation}
where we have $\xi_{min}=E[|b_{0}(i)-\mathbf{\tilde{w}}^{H}_{opt}\mathbf{r}(i)|^2]$ and $e_{a}(i)=\mbox{\boldmath$\varepsilon$}^{H}(i)\mathbf{r}(i)$ which denotes the error in    the beamformer coefficients $\mathbf{\tilde{w}}(i)$ via the \textit{a priori} estimation error.

When $i$ becomes a large number, since $\mathbf{\tilde{w}}(i)\rightarrow \mathbf{\tilde{w}}_{opt}$ and $E[\mbox{\boldmath$\varepsilon$}(i)]\rightarrow 0$ we have the steady-state MSE
\begin{equation}
\lim_{i\rightarrow \infty} \xi(i)= \xi_{min} + \lim_{i\rightarrow \infty} E[|e_{a}(i)|^2].
\end{equation}
Then, we define the steady-state excess MSE:
\begin{equation}
\xi_{ex}=\lim_{i\rightarrow \infty} E[|e_{a}(i)|^2].
\end{equation}
In the following, we derive  the expression for  $\xi_{ex}$.
Based on the energy conservation principle \cite{sayed1}-\cite{yousef},
in the steady state we have the energy preserving equation which is given as follows
\begin{equation}
E\big[\bar{\mu}(i)|e_{a}(i)|^2\big]=E\big[\bar{\mu}(i)\big|e_{a}(i)-\frac{\mu(i)}{\bar{\mu}(i)}F^{*}_{e}(i)\big|^2\big],\label{eq:energypre10}
\end{equation}
where $\bar{\mu}(i)=1/||\mathbf{B}\mathbf{r}(i)||^2$, $F^{*}_{e}(i)=\frac{e_{a}(i)-e_{p}(i)}{\mu(i) ||\mathbf{B}\mathbf{r}(i)||^2}$, $e_{p}(i)=\mbox{\boldmath$\varepsilon$}^{H}(i+1)\mathbf{r}(i)$ and $y(i)=\mathbf{\tilde{w}}^{H}(i)\mathbf{r}(i)=(\mathbf{\tilde{w}}_{opt}-\mbox{\boldmath$\varepsilon$}(i))^{H}\mathbf{r}(i)=\mathbf{\tilde{w}}^{H}_{opt}\mathbf{r}(i)-e_{a}(i)=b_{0}(i)+\bar{I}(i)+\bar{n}(i)-e_{a}(i)$,
where $\bar{I}(i)$ and $\bar{n}(i)$ denote the residual  interference and the residual noise, respectively,  as the output components of the optimum beamformer.

By expanding the right hand side (RHS) of (\ref{eq:energypre10}), we have
\begin{equation}
\begin{split}
E[\mu(i)]&E[e^{*}_{a}(i)y(i)(1-|y(i)|^2)]\\&\quad+E[\mu(i)]E[e_{a}(i)y^{*}(i)(1-|y(i)|^2)]\\&\quad=E[\mu^{2}(i)]E[||\mathbf{B}\mathbf{r}(i)||^2\underbrace{|y(i)|^2(1-|y(i)|^2)^2}_{|F_{e}(i)|^2}]. \label{eq:energypre11}
\end{split}
\end{equation}
Based on the analytical works in \cite{yousef} and \cite{whitehead}, we also make the following assumptions:
 \begin{enumerate}
\item In the steady state, the quantities $\{b_{0}(i), \bar{I}(i), \bar{n}(i), e_{a}(i) \}$ are zero-mean random variables, and they are mutually independent. The residual
 interference and the residual noise are Gaussian random variables.

\item In the steady state, $||\mathbf{B}\mathbf{r}(i)||^2$ and $|F_{e}(i)|^2$ are uncorrelated.

\item We have $E[b^{2l}_{k}(i)]=1$ for any positive integer $l$.
\end{enumerate}
Further, by employing the assumptions and substituting $y(i)=b_{0}(i)+\bar{I}(i)+\bar{n}(i)-e_{a}(i)$ into (\ref{eq:energypre11}) we have
\begin{equation}
\begin{split}
E[&\mu^2(i)]E[||\mathbf{B}\mathbf{r}(i)||^2]K_{1}E[|e_{a}(i)|^2]
\\&\quad+3E[\mu^2(i)]E[||\mathbf{B}\mathbf{r}(i)||^2]\sigma^{2}_{I}E[|e_{a}(i)|^4]\\&\quad
+3E[\mu^2(i)]E[||\mathbf{B}\mathbf{r}(i)||^2]\sigma^{2}_{n}E[|e_{a}(i)|^4]
\\&\quad+E[\mu^2(i)]E[||\mathbf{B}\mathbf{r}(i)||^2]E[|e_{a}(i)|^4]\\&\quad
+E[\mu^2(i)]E[||\mathbf{B}\mathbf{r}(i)||^2]K_{2}
\\&\quad+E[\mu^2(i)]E[||\mathbf{B}\mathbf{r}(i)||^2]E[|e_{a}(i)|^6]
\\&=2E[\mu(i)]\big(\sigma^{2}_{I}E[|e_{a}(i)|^2]+\sigma^{2}_{v}E[|e_{a}(i)|^2]+E[|e_{a}(i)|^4]  \big),
\end{split}
\end{equation}
where $K_{1}=3+3\sigma^{4}_{I}+6\sigma^{2}_{I}\sigma^{2}_{v}+3\sigma^{4}_{v}$,  $K_{2}=\sigma^{6}_{v}+3\sigma^{2}_{I}\sigma^{4}_{v}+3\sigma^{4}_{I}\sigma^{2}_{v}+\sigma^{6}_{I}+\sigma^{4}_{v}+2\sigma^{2}_{I}\sigma^{2}_{v}+\sigma^{4}_{I}+4\sigma^{2}_{v}+2\sigma^{2}_{I}+2$, $\sigma^{2}_{I}=E[\bar{I}^{2}(i)]$, and $\sigma^{2}_{v}=E[\bar{n}^{2}(i)]$.
When the filter works in the steady state, namely, $i$ becomes a large number, we assume $E[\bar{I}^{2l}(i)]=\big( E[\bar{I}^2(i)] \big)^{l}=\sigma^{2l}_{I}$ and $E[\bar{n}^{2l}(i)]=\big( E[\bar{n}^2(i)] \big)^{l}=\sigma^{2l}_{v}$.

Since the high power terms $E[|e_{a}(i)|^4]$ and $E[|e_{a}(i)|^6]$ can be neglected, we obtain the excess MSE as follows,
\begin{equation}
\begin{split}
\xi_{ex}&=E[|e_{a}(i)|^2]
\\&=\frac{E[\mu^{2}(\infty)]E[||\mathbf{B}\mathbf{r}(i)||^2]K_{2}}{2E[\mu(\infty)](\sigma^{2}_{I}+\sigma^{2}_{v})-E[\mu^{2}(\infty)]E[||\mathbf{B}\mathbf{r}(i)||^2]K_{1}}.
\end{split}
\end{equation}

We assume that the power of  residual  interference  at the output of the optimum beamformer is
significantly lower than the output noise power, namely, $\sigma^{2}_{I}\ll \sigma^{2}_{v}$. Thus, we can simplify the expression for the excess MSE as follows
\begin{equation}
\begin{split}
\xi_{ex}&=E[|e_{a}(i)|^2]
\\&\approx\frac{E[\mu^{2}(\infty)]E[||\mathbf{B}\mathbf{r}(i)||^2](\sigma^{6}_{v}+\sigma^{4}_{v}+4\sigma^{2}_{v}+2)}{2E[\mu(\infty)]\sigma^{2}_{v}-E[\mu^{2}(\infty)]E[||\mathbf{B}\mathbf{r}(i)||^2](3+3\sigma^{4}_{v})}.\label{eq:steadystateMSE1}
\end{split}
\end{equation}

In order to compute the final excess MSE, we also need to derive the steady-state first order and second order statistical expressions for the variable step-size values.
By employing the methodology in \cite{werner11}, when $i$ becomes a large number, we obtain the following:
\begin{equation}
E[\mu(\infty)]=E[\gamma(i)] \mathcal{P}+\frac{(1-\mathcal{P})}{E[\gamma(i)]E[||\mathbf{B}\mathbf{r}(i)||^2]},\label{eq:steadystateMSE2}
\end{equation}
\begin{equation}
E[\mu^{2}(\infty)]=E[\gamma(i)] \mathcal{P}+\frac{(1-\mathcal{P})}{E[\gamma(i)]E[||\mathbf{B}\mathbf{r}(i)||^4]},\label{eq:steadystateMSE3}
\end{equation}
where $\mathcal{P}$ denotes the probability of update at the steady-state, which is given by
\begin{equation}
\begin{split}
\mathcal{P}&=\textit{Pr}\{E[|e(i)|^2]>E[|\gamma(i)|^2]\}\approx \textit{Pr}\{|e(i)|>E[\gamma(i)]\}\\& \approx 2Q\bigg(\frac{E[\gamma(i)]}{\sigma_{v}} \bigg),
\end{split}
\end{equation}
where $i$ is a very large number,
   $\textit{Pr}\{.\}$ denotes the probability, and $Q(x)$ is the complementary Gaussian cumulative distribution function \cite{rappaport} which is given by $Q(x)=\int^{\infty}_{x} \frac{1}{\sqrt{2\pi}}e^{\frac{-t^{2}}{2}} dt$.
The expression of $E[\gamma(i)]$ for the PDB  scheme at the steady state can be derived based on (\ref{eq:bound1}) and is given by
\begin{equation}
E[\gamma(i)]=\sqrt{\lambda}  {\sigma}_{n} ||\mathbf{\tilde{w}}_{opt} ||.\label{eq:steadystateMSE4}
\end{equation}
By following the same approach and using (\ref{eq:bound3}), we have the expression of $E[\gamma(i)]$ for the PIDB  scheme at the steady state:
\begin{equation}
\begin{split}
E[\gamma(i)]
 \approx \sqrt{\psi}\sqrt{E[ \nu(i)]}+\sqrt{\lambda}{\sigma}_{n} ||\mathbf{\tilde{w}}_{opt} ||.\label{eq:steadystateMSE5}
\end{split}
\end{equation}
From (\ref{eq:bound2}), when $i$ becomes a large number, we have $E[\nu(i)]=E[|\mathbf{w}^{H}_{opt}\mathbf{B}\mathbf{r}(i)|^{2}]=\mathbf{w}^{H}_{opt}\mathbf{B}\big(\sum^{q-1}_{k=1}\mathbf{a}(\theta_{k})\mathbf{a}^{H}(\theta_{k})+\sigma^{2}_{n}\mathbf{I} \big)\mathbf{B}^{H}\mathbf{w}_{opt}$. In the simulations, we will show  the effectiveness of our derivation and approximation.


\section{Simulations}
\label{Section6:simulations}

In this section, we evaluate the performance of the proposed set-membership adaptive blind beamforming algorithms  and compare them with the existing adaptive blind beamforming algorithms including the conventional adaptive SG beamforming algorithms based on the CM and MV criteria with GSC structures.
We carried out simulations to assess the convergence performance of signal-to-interference-plus-noise ratio (SINR) against the number of snapshots.
%
In the simulations, we assume that there is one desired user in the system and the related DOA is known by the receiver. Simulations are performed with a ULA containing $m=16$ sensor
elements with half-wavelength inter-element spacing. The DOAs are randomly generated with uniform random variables between $0$ and $180$ degrees
for each experiment. The results are averaged by $1000$ runs. We consider the binary phase shift keying (BPSK) modulation and set $v=1$.

\begin{figure}[!hhh]
\centering \scalebox{0.58}{\includegraphics{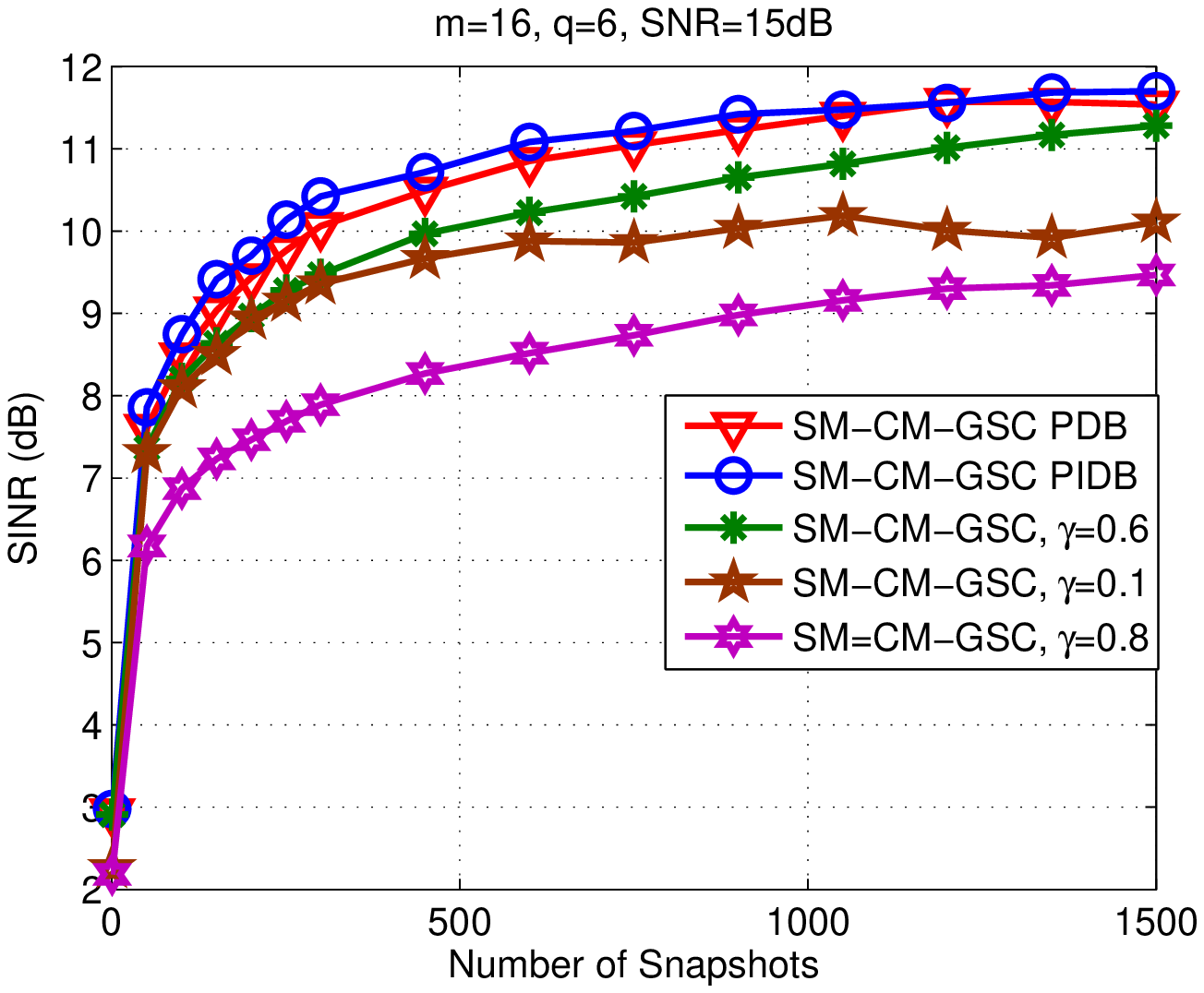}}
\caption{Output SINR versus the number of snapshots. The number of users is $q=6$. The update rates for the proposed SM-CM-GSC algorithms with PIDB and PDB  are $22.4 \%$ and $26.5 \%$, respectively.
}
\label{fig:fig1}
\end{figure}

In Fig. \ref{fig:fig1}, we compare the proposed SM-CM-GSC adaptive beamforming algorithm with fixed bounds and that with time-varying bounds. We consider a scenario with $q=6$ users with the same power level in the system. The input signal-to-noise ratio (SNR) is $15$ dB. The initial values of the weight vector is given by $\mathbf{w}(0)=[1,0,\ldots,0]^{T}$. The coefficients for the PDB and PIDB schemes are given by $\rho=0.98$, $\lambda=2$, $\psi=0.003$, $\gamma(0)=0$ and $v(0)=0$. For the fixed bound scheme, we set $\gamma=0.1$, $\gamma=0.6$ and $\gamma=0.8$ to test the performance.
The simulation results firstly illustrate that the SM-CM-GSC adaptive beamforming algorithm with $\gamma=0.6$ provides a better performance compared to the other choices for the fixed bound. Secondly, we can see that the proposed beamforming algorithms with time-varying bound schemes outperform the beamforming algorithms with fixed bound schemes. Moreover, the performance with the PIDB scheme is slightly better than the performance  with the PDB scheme. Due to the data-selective update feature the SM-CM-GSC adaptive algorithms with $\gamma=0.1$, $\gamma=0.6$ and $\gamma=0.8$ can provide $72.9\%$, $27.6\%$  and $14.8\%$ update rates, respectively. The proposed beamforming algorithms with the PIDB and PDB schemes have $22.4\%$ and $26.5\%$ update rates, respectively.

\begin{figure}[!hhh]
\centering \scalebox{0.58}{\includegraphics{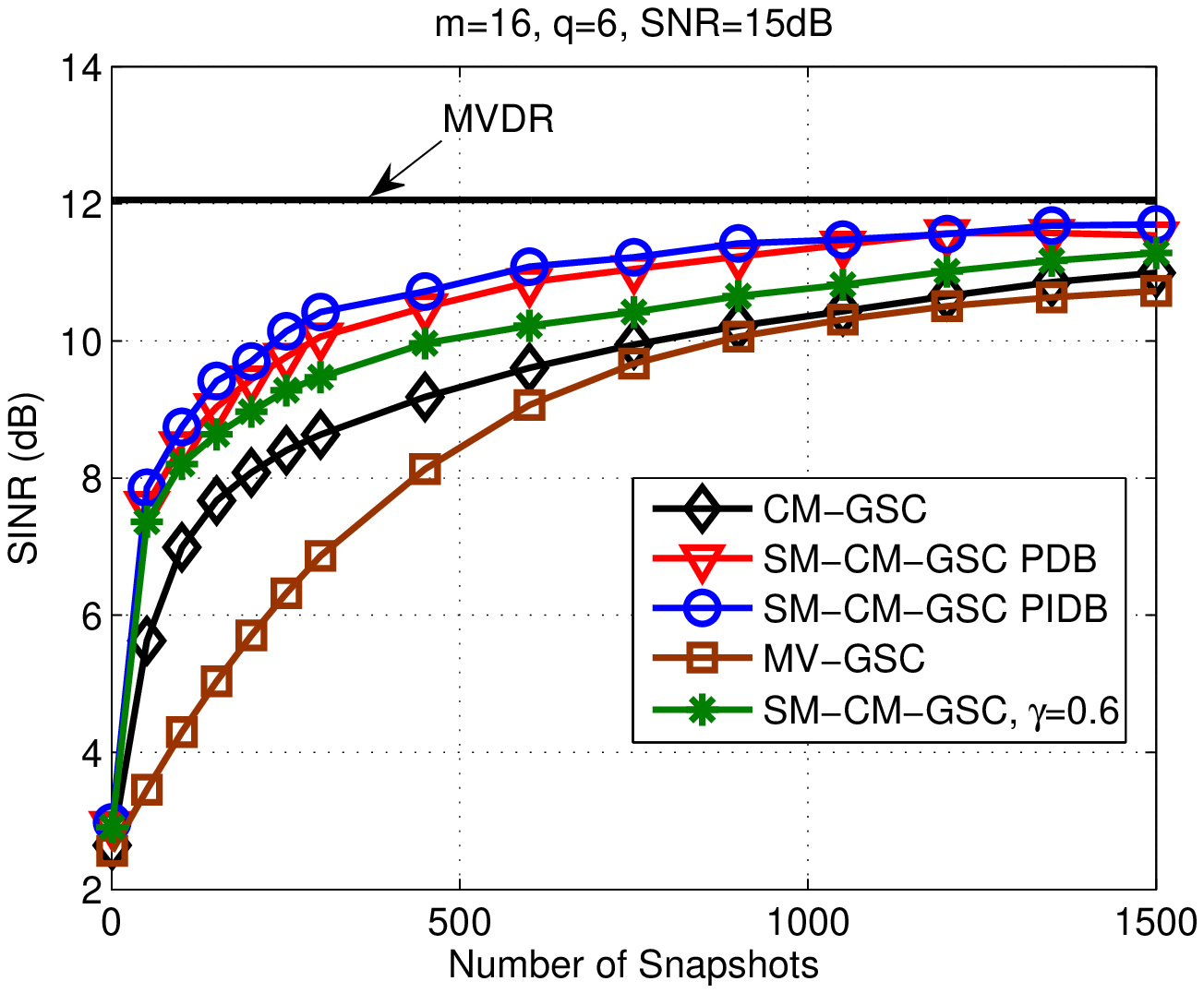}}
\caption{Output SINR versus the number of snapshots. The number of users is $q=6$. The update rates for the proposed SM-CM-GSC algorithms with PIDB and PDB  are $22.4 \%$ and $26.5 \%$, respectively.
}
\label{fig:fig2}
\end{figure}

\begin{figure}[!hhh]
\centering \scalebox{0.58}{\includegraphics{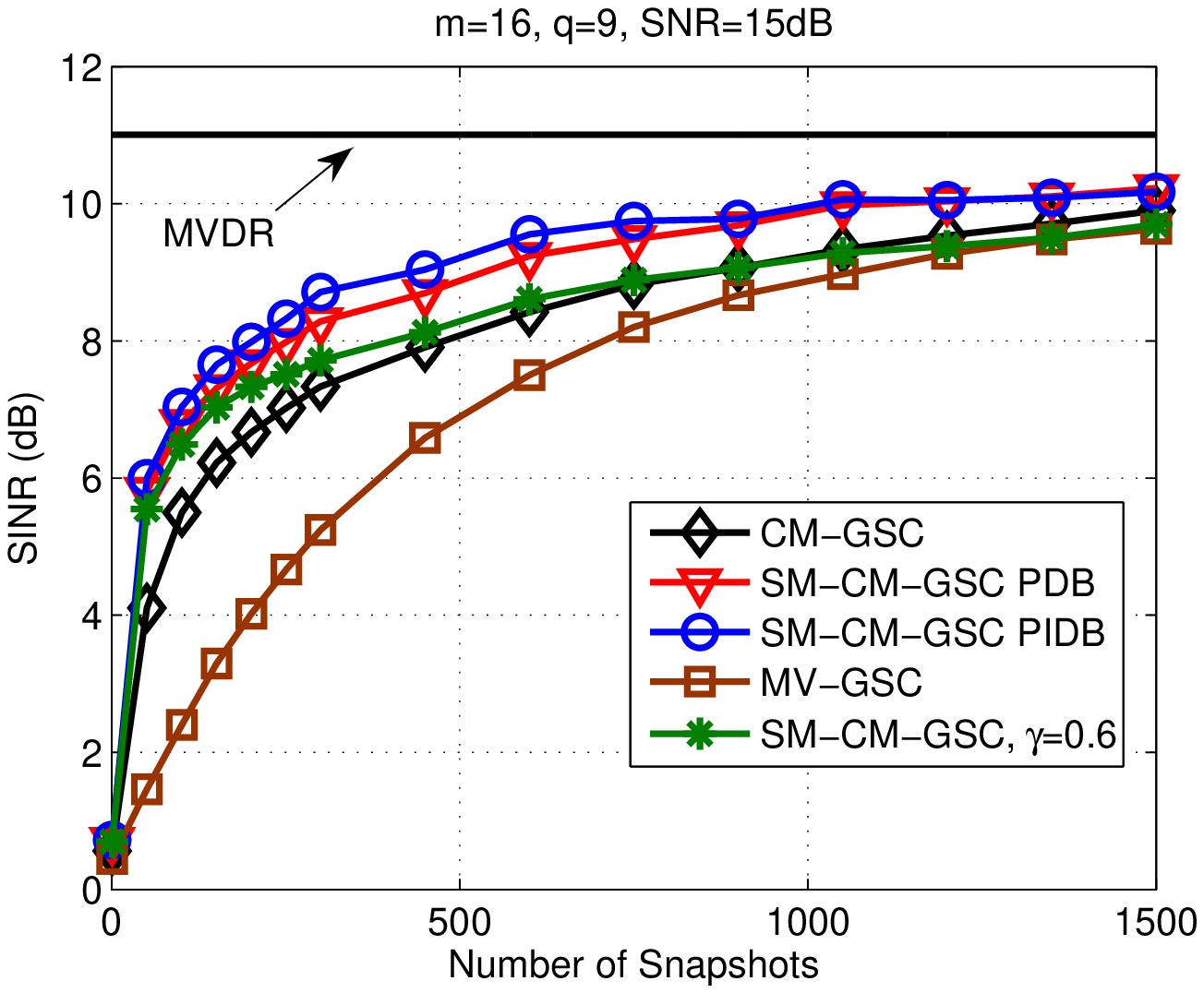}}
\caption{Output SINR versus the number of snapshots. The number of users is $q=9$. The update rates for the proposed SM-CM-GSC algorithms with PIDB and PDB  are $23 \%$ and $27 \%$, respectively.
}
\label{fig:fig3}
\end{figure}

Fig. \ref{fig:fig2} and Fig. \ref{fig:fig3} indicate the SINR convergence performance versus the number of snapshots for the proposed SM-CM-GSC adaptive algorithms and the conventional adaptive beamforming algorithms in the presence of different number of users. The input SNR is $15$ dB. The number of users corresponding to the results in Fig. \ref{fig:fig2} and Fig. \ref{fig:fig3} are $q=6$ and $q=9$, respectively. The coefficients of the proposed adaptive beamforming algorithms with time-varying bound schemes are well tuned as the simulations of Fig. \ref{fig:fig1}. The fixed bound of the SM-CM-GSC algorithm is $\gamma=0.6$. The step-size values of the conventional SG adaptive  CM-GSC and MV-GSC algorithms are tuned as $\mu=0.005$.
We note that all the parameters for the analyzed algorithms are optimized based on simulations.
From the results, we can see that the proposed SM-CM-GSC adaptive beamforming algorithms with the PIDB and PDB schemes achieve the best convergence performance. While they only require around $20\%$ of the time for filter parameter updates and can save significant computational resources. The performance of the minimum variance distortionless response (MVDR) beamforming solution is given as a reference.

\begin{figure}[!hhh]
\centering \scalebox{0.58}{\includegraphics{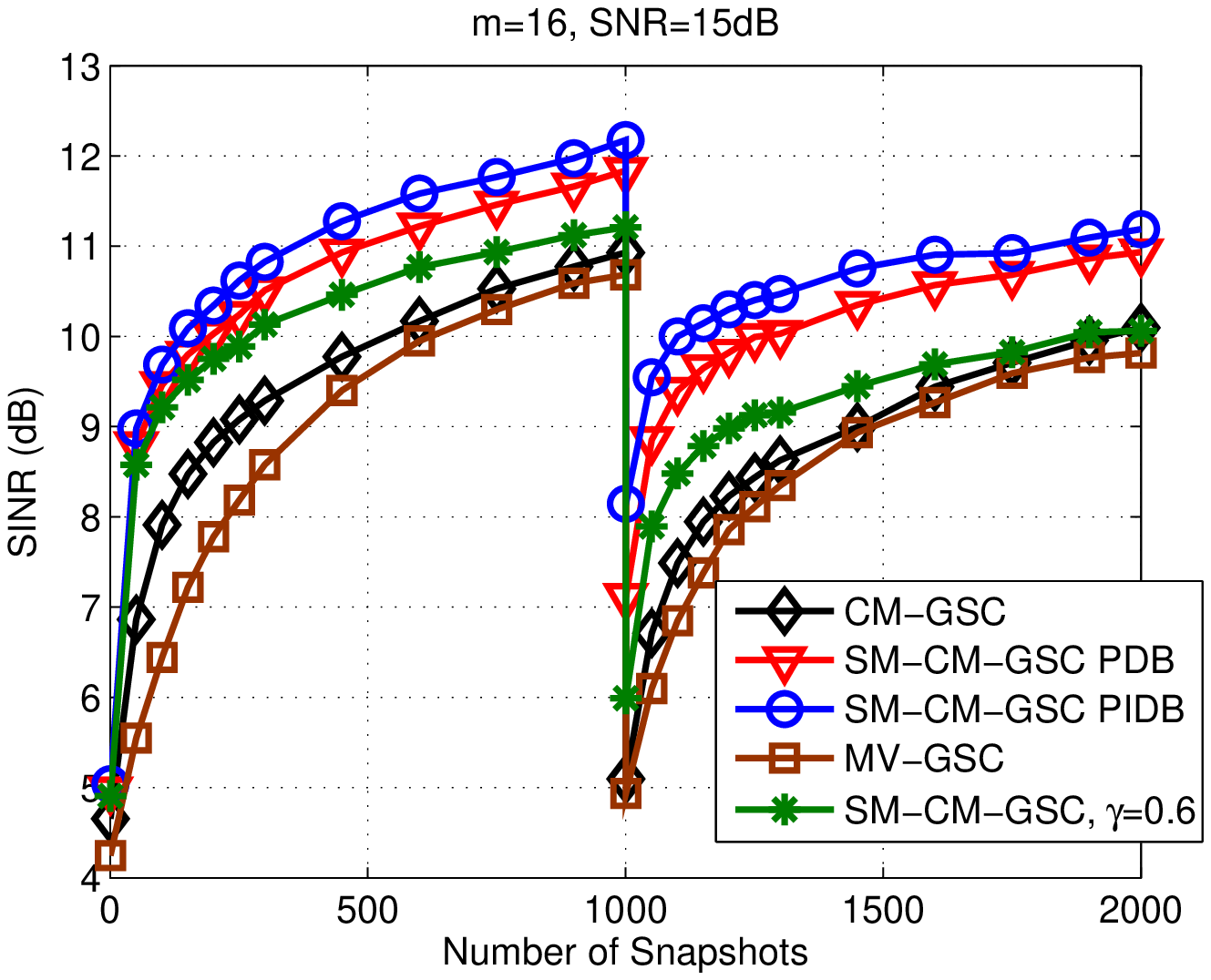}}
\caption{Output SINR versus the number of snapshots in a nonstationary scenario. SNR=$15$ dB.
The update rates for the proposed SM-CM-GSC algorithms with PIDB and PDB  are $25 \%$ and $30 \%$, respectively.
}
\label{fig:fig4}
\end{figure}

Fig. \ref{fig:fig4} shows the convergence performance in a nonstationary scenario.
The
system starts with four users including one high-power level
interferer with $3$ dB above the desired user. At $1000$ snapshots, three interferers including one user operating at $3$ dB above  the desired user's power level
 enter the system.
 From the results, we can see that the proposed SM-CM-GSC adaptive algorithm with the PIDB  scheme  achieves the best performance, followed by
 the SM-CM-GSC adaptive algorithm with the PDB scheme,  the SM-CM-GSC adaptive algorithm with a fixed bound, the CM-GSC adaptive algorithm and
the MV-GSC algorithm.
The proposed algorithm is more robust to dynamic scenarios compared to the conventional SG-based algorithms.
The  SNR is $15$dB. We set $\rho=0.98$, $\lambda=2$, $\psi=0.003$, $\gamma(0)=0$ and $v(0)=0$. The fixed bound  is  chosen as $\gamma=0.6$.

%

\begin{figure}[!hhh]
\centering \scalebox{0.58}{\includegraphics{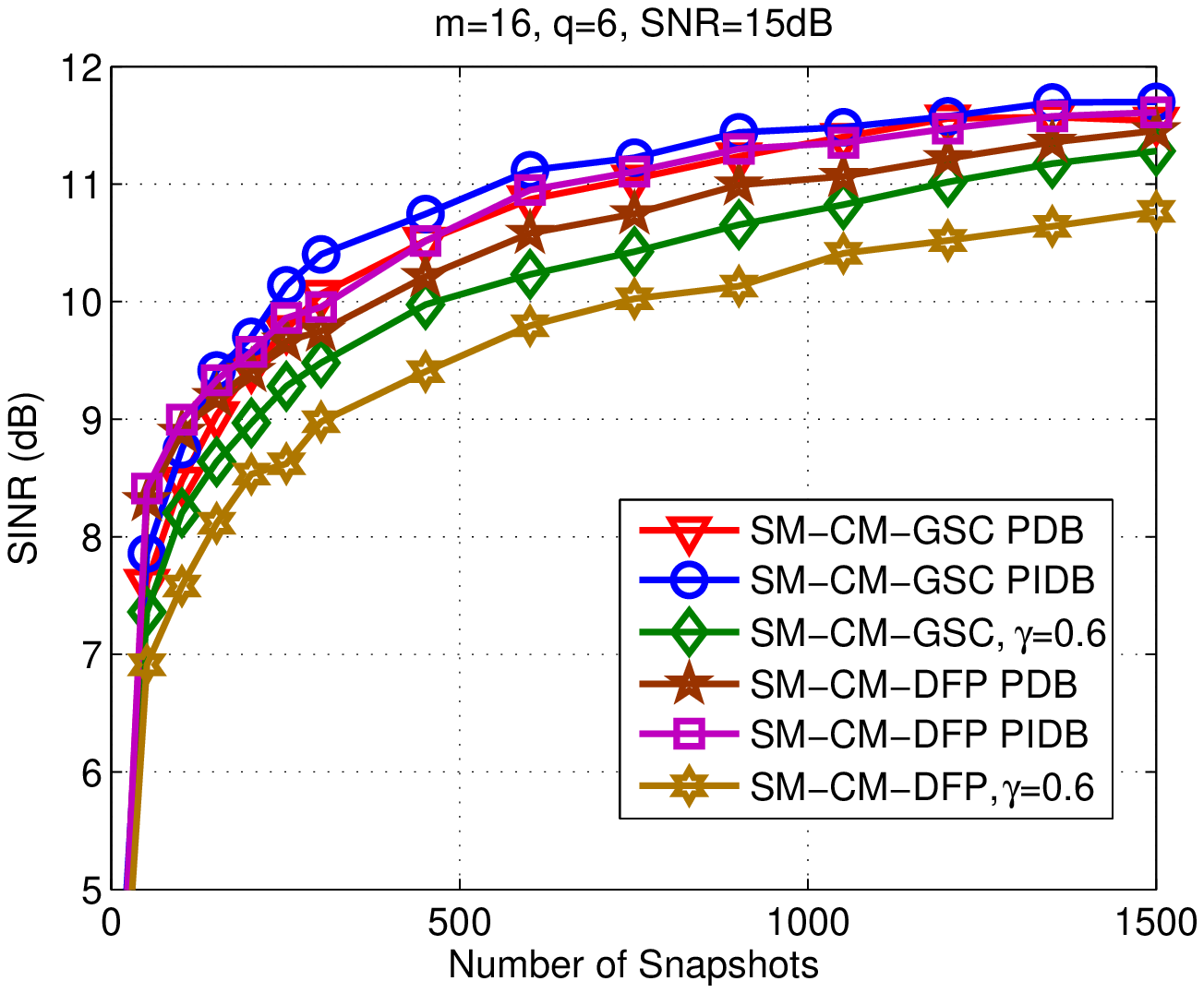}}
\caption{Output SINR versus the number of snapshots. SNR=$15$ dB. The number of users is $q=6$.
The update rates for the proposed adaptive SM-CM-GSC algorithms with PIDB and PDB  are $22.4 \%$ and $26.5 \%$.
The update rates for the adaptive  SM-CM-DFP algorithms with PIDB and PDB  are $18.7 \%$ and $25.4 \%$.
The update rates for the adaptive SM-CM-GSC and SM-CM-DFP algorithms with a fixed error bound are $27.6 \%$ and $27 \%$, respectively.
}
\label{fig:fig5}
\end{figure}

In the next simulation, we investigate the set-membership adaptive CM beamformers employing the DFP structure  \cite{sm-ccm}, and  compare them with the proposed set-membership adaptive CM beamforming algorithms with the GSC structure. In particular, we investigate the SM-CM beamforming algorithms with the PIDB and PDB schemes for both DFP and GSC structures.
 The results as shown in Fig. \ref{fig:fig5} illustrate that the convergence performance of our proposed SM-CM-GSC algorithm with the PIDB scheme is slightly better than the performance of the SM-CM-DFP algorithm with the PIDB scheme. While the SM-CM-GSC algorithm with the PDB scheme outperforms the SM-CM-DFP algorithm with PDB scheme. In the experiment, the number of users is $q=6$ and the input SNR is $15$ dB. The coefficients of the GSC and DFP based beamforming algorithms were well tuned as the ones in the previous simulations.

\begin{figure}[!hhh]
\centering \scalebox{0.46}{\includegraphics{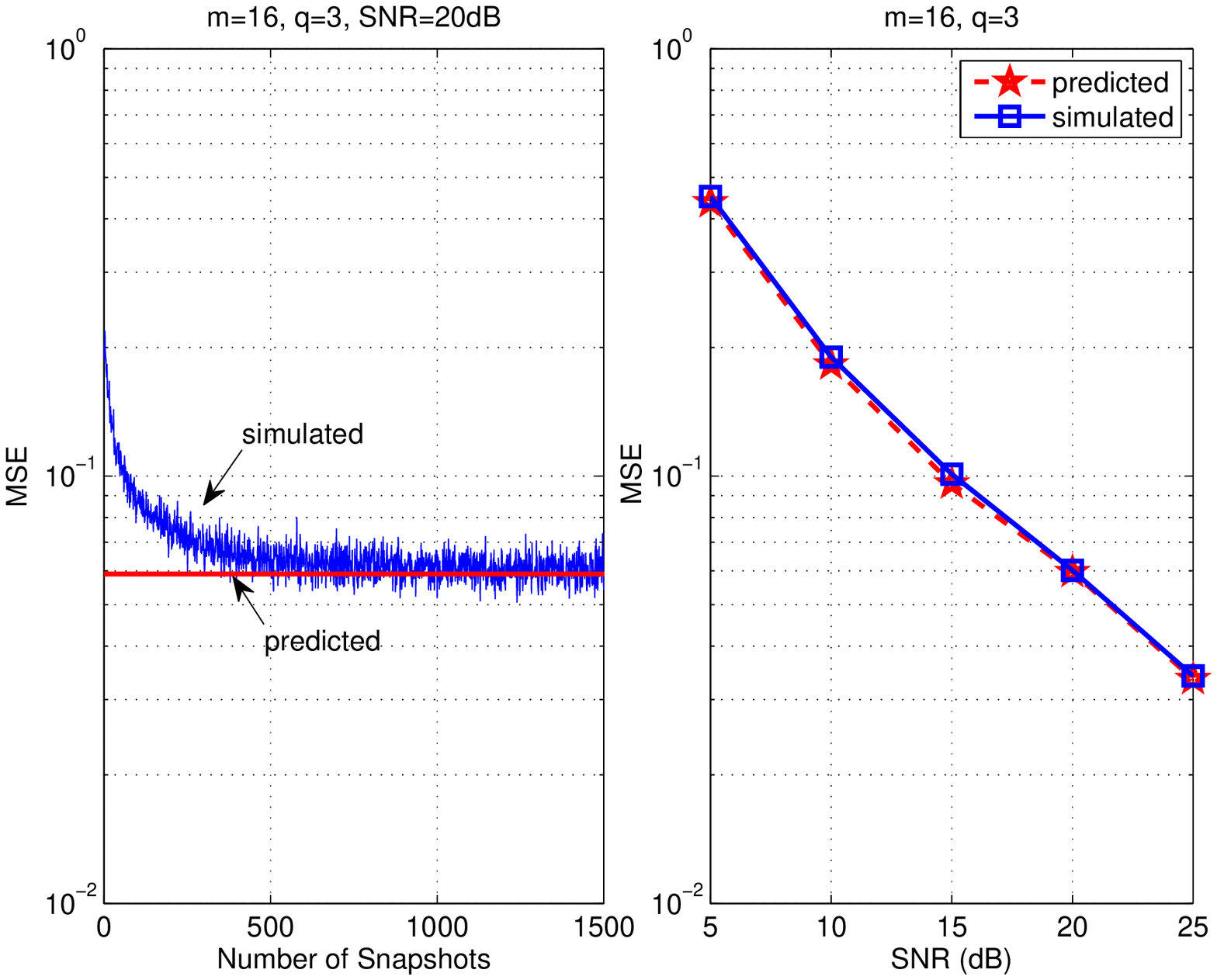}}
\caption{Analytical MSE versus simulated performance for convergence analysis of the proposed SM adaptive beamforming algorithm with PIDB scheme. (a) The number of users is $q=3$, SNR=$20$ dB. (b) The number of users is $q=3$. }
\label{fig:fig6}
\end{figure}

We consider the convergence  analysis of the
proposed adaptive SM-CM-GSC beamformer with the PIDB scheme. The
steady-state MSE between the desired and the estimated signal
obtained through simulation is compared with the steady-state
MSE computed via the expressions derived in Section \ref{Section5:analysis}.
We verify that the analytical results (\ref{eq:steadystateMSE1}), (\ref{eq:steadystateMSE2}), (\ref{eq:steadystateMSE3}) and (\ref{eq:steadystateMSE5}) are able
to predict the steady-state MSE.
As the work proposed in \cite{whitehead}, we use a scaled version of the
Wiener filter to approximate the optimum  CM-GSC solution.
In this simulation of convergence
analysis, we assume that three users having the same
power level operate in the system. By comparing the curves in Fig. \ref{fig:fig6}(a), it can be
seen that as the number of snapshots increases and the
simulated MSE converges to the analytical result, showing the
usefulness of our analysis and assumptions. Fig. \ref{fig:fig6}(b) shows the MSE performance
versus the desired user¡¯s SNR and a comparison between the
steady-state analysis and simulation results. The simulation and
analysis results agree well with each other.

\begin{figure}[!hhh]
\centering \scalebox{0.46}{\includegraphics{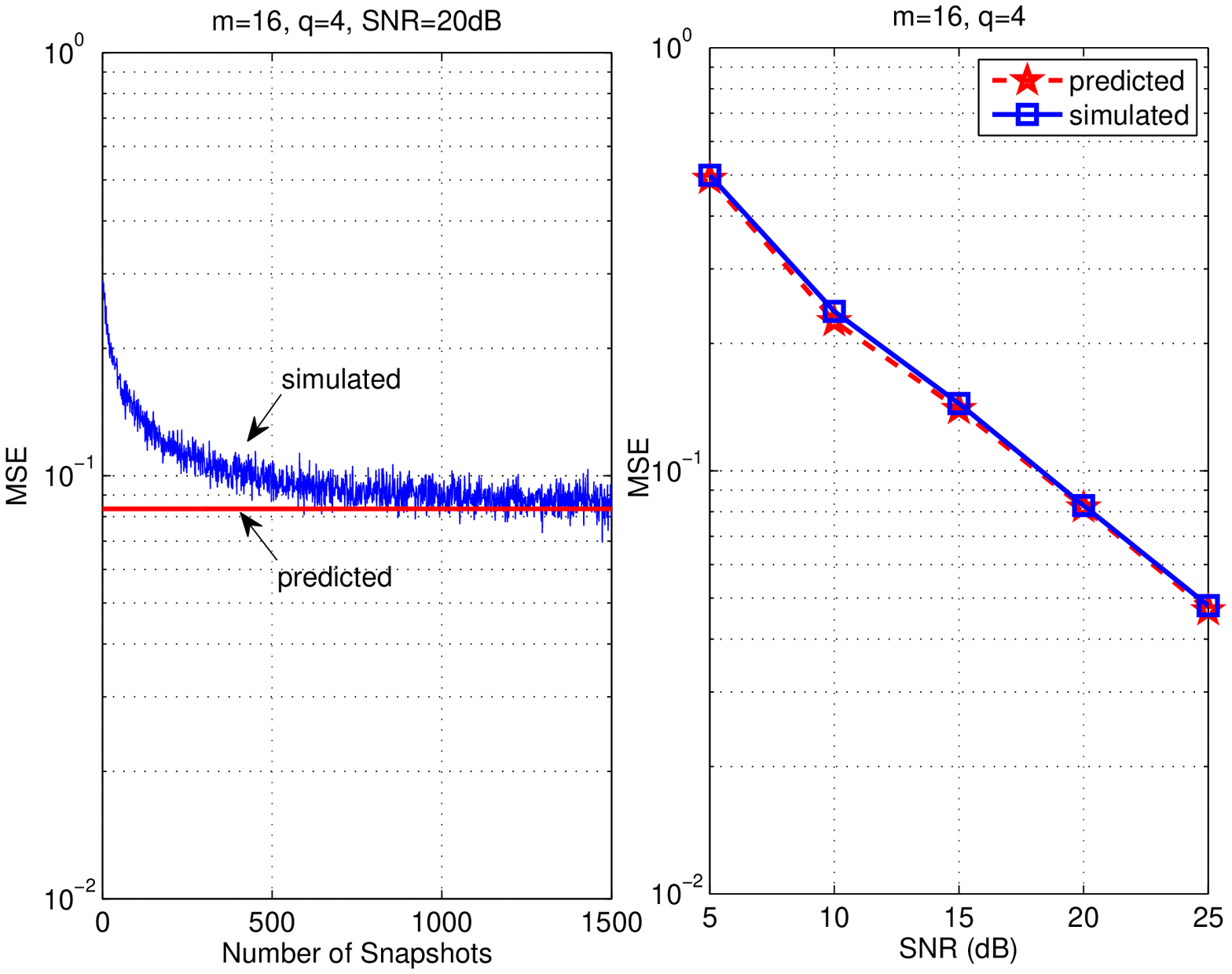}}
\caption{Analytical MSE versus simulated performance for convergence analysis of the proposed SM adaptive beamforming algorithm with PDB scheme.  (a) The number of users is $q=4$, SNR=$20$ dB. (b) The number of users is $q=4$.}
\label{fig:fig7}
\end{figure}

In the final simulation results, we discuss the convergence analysis of the proposed
SM-CM-GSC beamforming algorithm with the PDB scheme. Here, we verify that the analytical results (\ref{eq:steadystateMSE1}), (\ref{eq:steadystateMSE2}), (\ref{eq:steadystateMSE3}) and (\ref{eq:steadystateMSE4}) are able to provide an accurate prediction of the steady-state MSE. In this simulation, we assume that four
users operate with the same power level in the system.  Fig. \ref{fig:fig7}(a) indicates that as the number of snapshots
increases, the simulated MSE converges to the analytical
result, showing the usefulness of our convergence analysis for the PDB scheme. Fig. \ref{fig:fig7}(b) shows the effect that
the desired user¡¯s SNR has on the MSE. We also can see that
the simulation and analysis results agree well with each other.

\section{Conclusion}
\label{Section7:Conclusions}
In this paper, we have proposed an adaptive blind set-membership beamforming algorithm with a GSC structure using the CM criterion.
We have developed a SG-type algorithm based on the concept of SM filtering for  adaptive implementation.
We updated the filter weights only if the  constraint cannot be satisfied. Moreover, two schemes of time-varying bounds have been proposed to blind beamforming with a GSC structure. We have also incorporated parameter and interference dependence to characterize  the environment for improving the tracking performance of the proposed algorithm.
For the proposed adaptive algorithm, we have investigated the convergence and
derived expressions to predict the steady-state MSE.
Simulation results have shown that the proposed blind SM beamforming algorithm with dynamic bounds achieves superior performance to previously reported methods at a reduced update rate.

\appendix

\section{Analysis of the Optimization Problem}
\label{section8:appendixa}

In this part, we discuss the convexity of the cost function which is expressed in (\ref{eq:cm1}).
Without loss of generality, we assume that user $0$ is the desired user.
We rewrite the cost function as follows,
\begin{equation}
\begin{split}
J_{CM}&=E\Big[ \big( | \mathbf{\tilde{w}}^{H}(i)\mathbf{r}(i) |^2-1  \big)^{2} \Big]\\&
=E\Big[ \big(|y(i)|^2 -1 \big)^2 \Big]\\&
=E\big[ |y(i)|^4\big]-2E\big[ |y(i)|^2\big]+1.
\end{split}
\end{equation}
 Let us define $z_{1}(i)=\mathbf{\tilde{w}}^{H}(i)\mathbf{A}(\mathbf{\mbox{\boldmath$\theta$}})\mathbf{b}(i)=\mathbf{s}^{H}\mathbf{b}(i)$ and $z_{2}(i)=\mathbf{\tilde{w}}^{H}(i)\mathbf{n}(i)$, where $\mathbf{s}=[s_{0},\ldots, s_{q-1}]^{T}$ and $s_{k}= \mathbf{a}^{H}(\theta_{k})\mathbf{\tilde{w}}(i)$, $k=0,\ldots, q-1$. By letting $D=s_{0}s^{*}_{0}=v^2$ and $\mathbf{\bar{s}}=[s_{1},\ldots,s_{q-1}]^{T}$,
 we obtain
\begin{equation}
J_{CM}=J_{1}(\mathbf{\bar{s}})+\sigma^{2}_{n}J_{2}\big(\mathbf{\tilde{w}}\big),
\end{equation}
where
\begin{equation}
\begin{split}
J_{1}(\mathbf{\bar{s}})&=2(D+\mathbf{\bar{s}}^{H}\mathbf{\bar{s}})^2-\big(D^2+\sum^{q-1}_{k=1}s^{4}_{k}  \big)
-2(D+\mathbf{\bar{s}}^{H}\mathbf{\bar{s}})+1, \label{eq:costfunction1}
\end{split}
\end{equation}
\begin{equation}
\begin{split}
J_{2}\big(\mathbf{\tilde{w}}\big)=\big(4(D+\mathbf{\bar{s}}^{H}\mathbf{\bar{s}})-2+3\sigma^2_{n}\mathbf{\tilde{w}}^{H}\mathbf{\tilde{w}} \big)\mathbf{\tilde{w}}^{H}\mathbf{\tilde{w}}. \label{eq:costfunction2}
\end{split}
\end{equation}
In order to evaluate the convexity of $J_{CM}$, we compute its Hessian matrix by using the rule $\mathbf{M}=\frac{\partial}{\partial\mathbf{\tilde{w}}^{H}}\frac{\partial J_{CM}}{\partial\mathbf{\tilde{w}}} $ which yields $\mathbf{M}=\mathbf{M}_{1}+\sigma^2_{n}\mathbf{M}_{2}$, where
\begin{equation}
\begin{split}
\mathbf{M}_{1}&=4\mathbf{\bar{A}}\big[(D-1/2)\mathbf{I}+\mathbf{\bar{s}}^{H}\mathbf{\bar{s}}\mathbf{I}+\mathbf{\bar{s}}\mathbf{\bar{s}}^{H}- \textrm{diag} \big(|s_{1}|^2, \ldots, |s_{q-1}|^2  \big)  \big] \mathbf{\bar{A}}^{T},\label{eq:cm2}
\end{split}
\end{equation}
\begin{equation}
\begin{split}
\mathbf{M}_{2}&=(4D-2)\mathbf{I}+6\sigma^2_{n}\big(\mathbf{\tilde{w}}^{H}\mathbf{\tilde{w}}\mathbf{I}+\mathbf{\tilde{w}}\mathbf{\tilde{w}}^{H} \big)+4\big( \mathbf{\tilde{w}}^{H}\mathbf{\bar{A}}\mathbf{\bar{A}}^{H} \mathbf{\tilde{w}} \mathbf{I}+ (\mathbf{\bar{A}}\mathbf{\bar{A}}^{H})^{T}\mathbf{\tilde{w}}^{H}\mathbf{\tilde{w}}\\&\quad+(\mathbf{\tilde{w}}\mathbf{\tilde{w}}^{H}\mathbf{\bar{A}}\mathbf{\bar{A}}^{H})^{T}+
(\mathbf{\tilde{w}}^{H}\mathbf{\bar{A}}\mathbf{\bar{A}}^{H}\mathbf{\tilde{w}})^{T}  \big),\label{eq:cm3}
\end{split}
\end{equation}
where $\mathbf{\bar{A}}=[\mathbf{a}(\theta_{1}), \ldots, \mathbf{a}(\theta_{q-1})]$.

The matrix $\mathbf{M}$ is positive definite if $\mathbf{\mbox{\boldmath$\alpha$}}^{H}\mathbf{M}\mathbf{\mbox{\boldmath$\alpha$}}>0$ for any nonzero $(q-1)\times 1$ vector $\mathbf{\mbox{\boldmath$\alpha$}}$.  The second, third and fourth terms for  $\mathbf{M}_{1}$ in (\ref{eq:cm2}) yield the positive definite matrix $4\big(\mathbf{\bar{s}}\mathbf{\bar{s}}^{H}+\textrm{diag}\big(|s_{1}|^2, \ldots, |s_{q-1}|^2 \big)  \big)$, while the first term provides the condition $D=v^2\geq 1/2$ that ensures the convexity of $J_{1}(\mathbf{\bar{s}})$. Therefore, when $\sigma^{2}_{n}=0$, the function $J_{CM}$ is  convex. Since $J_{CM}$ is continuous in terms of $\sigma^{2}_{n}$, we may assume that the extrema of the cost function in noisy case can be deduced for small $\sigma^{2}_{n}$ by a slight perturbation of the noiseless extrema \cite{cxu21}. For the matrix $\mathbf{M}_{2}$ in (\ref{eq:cm3}), it is easily seen that we can select a sufficiently large value of $D$ such that $\mathbf{M}_{2}$ is positive definite in any bounded region. Recalling that $D=v^2$, we obtain that with properly selecting the constant $v$, $\mathbf{M}$ is positive definite in any bounded region, which results in the cost function $J_{CM}$ being strictly convex. The algorithm is then able to reach the global minima under these assumptions.

\section{Proof of (\ref{eq:costfunction1}) and (\ref{eq:costfunction2})}
\label{section9:appendixb}

We know that the cost function can be expressed as follows,
\begin{equation}
J_{CM}=E[|y(i)|^4]-2E[|y(i)|^2]+1,
\end{equation}
where
\begin{equation}
y(i)=z_{1}(i)+z_{2}(i)=\mathbf{s}^{H}\mathbf{b}(i)+\mathbf{\tilde{w}}^{H}(i)\mathbf{n}(i).
\end{equation}
In order to further investigate the cost function, we need to assess $E[|y(i)|^4]$ and $E[|y(i)|^2]$.
By assuming that the source signals and the complex Gaussian noise are independent and identically distributed, we have
\begin{equation}
E[|y(i)|^4]=E[|z_{1}(i)|^4]+E[|z_{2}(i)|^4]+4E[|z_{1}(i)|^2|z_{2}(i)|^2],
\end{equation}
\begin{equation}
E[|y(i)|^2]=E[|z_{1}(i)|^2]+E[|z_{2}(i)|^2].
\end{equation}
Because the source signal takes on the value $+1$ with probability $0.5$ or the value $-1$ with the same probability, we have
\begin{equation}
\begin{split}
E[|z_{1}(i)|^4]&=E[(\mathbf{s}^{H}\mathbf{b}\mathbf{b}^{H}\mathbf{s})^2]=E\bigg[\bigg(\sum^{q-1}_{i=0}\sum^{q-1}_{j=0}s^{*}_{i}b_{i}b_{j}s_{j}\bigg)^2\bigg]
\\&\quad=E\bigg[  \sum^{q-1}_{i=0} \sum^{q-1}_{j=0} \sum^{q-1}_{l=0} \sum^{q-1}_{n=0} b_{i}b_{j}b_{l}b_{n}s^{*}_{i}s_{j}s^{*}_{l}s_{n}  \bigg].
\end{split}
\end{equation}
Since $s$ and $b$ are independent, we have $E[b_{i}]=0$ for $\forall i$  and $E[b_{i}b_{j}]=0$ for $i\neq j$, the only non-zero terms in the sum arise when the product of four values of $b$ at various times can be grouped into two pairs, i.e., $E[b_{i}b_{j}b_{l}b_{n}]=E[b^{2}_{1}b^{2}_{2}]$ with $b_{1}\neq b_{2}$ permitted. Exactly three grouping are possible from a set of four that gives a particular two pairs with different time arguments for the two pairs. However, only one possibility has them all the same. Thus,
\begin{equation}
\begin{split}
E\big[|z_{1}(i)|^4 \big]&= \sum^{q-1}_{i=0}|s_{i}|^4+2\sum^{q-1}_{i=0}\sum^{q-1}_{l=0,\neq i}|s_{i}|^2|s_{l}|^2
+\sum^{q-1}_{i=0}\sum^{q-1}_{j=0,\neq i}s_{i}s^{*}_{j}s_{i}s^{*}_{j},\label{eq:appen1}
\end{split}
\end{equation}
because the $s$ are sequentially uncorrelated, $E[s_{i}s^{*}_{j}s_{i}s^{*}_{j}]=E[s_{i}s_{i}]E[s^{*}_{j}s^{*}_{j}]=0$ for $i\neq j$.
We use the observation to convert (\ref{eq:appen1}) to
\begin{equation}
E\big[|z_{1}(i)|^4 \big]=\sum^{q-1}_{i=0}|s_{i}|^4+2\sum^{q-1}_{i=0}\sum^{q-1}_{l=0,\neq i} |s_{i}|^2 |s_{l}|^2.
\end{equation}
Thus, we can obtain
\begin{equation}
J_{1}(\mathbf{\bar{s}})=2(\mathbf{s}^{H}\mathbf{s})^2-\sum^{q-1}_{k=0}s^{4}_{k}-2\mathbf{s}^{H}\mathbf{s}+1,
\end{equation}
\begin{equation}
J_{2}(\mathbf{\tilde{w}})=(4\mathbf{s}^{H}\mathbf{s}-2+3\sigma^2_{n}\mathbf{\tilde{w}}^{H}\mathbf{\tilde{w}})\mathbf{\tilde{w}}^{H}\mathbf{\tilde{w}}.
\end{equation}

\section{Derivation for (\ref{eq:smcmgsc2})}
\label{section10:appendixc}

 By imposing the condition to update whenever $e^2(i) > \gamma^2(i)$,
 we can obtain $\mu(i)$ in order to compute $\mathbf{w}(i+1)$ by projecting $\mathbf{w}(i)$ onto $\mathcal{H}_{i}$, i.e., the set of all $\mathbf{w}$ that satisfy:
\begin{equation}
\sqrt{1-\gamma(i)}\leq | (v\mathbf{a}(\theta_{0})- \mathbf{B}^{H}\mathbf{w} )^{H}\mathbf{r}(i)| \leq \sqrt{1+\gamma(i)}.
\end{equation}
The set comprises two parallel hyper-strips in the parameter space. For  case 1): $|y(i)|=| (v\mathbf{a}(\theta_{0})- \mathbf{B}^{H}\mathbf{w}(i) )^{H}\mathbf{r}(i)| > \sqrt{1+\gamma(i)}$, $\mathbf{w}(i)$ is closer to the hyperplanes defined by $|(v\mathbf{a}(\theta_{0})- \mathbf{B}^{H}\mathbf{w} )^{H}\mathbf{r}(i)|=\sqrt{1+\gamma(i)}$ than to the ones defined by $|(v\mathbf{a}(\theta_{0})- \mathbf{B}^{H}\mathbf{w} )^{H}\mathbf{r}(i)|=\sqrt{1-\gamma(i)}$.
By employing (\ref{eq:ccm3}) we have
 \begin{equation}
 \begin{split}
 |y(i)&+\mu(i)\big(\mathbf{B}\mathbf{r}(i)y^{*}(i)-|y(i)|^2\mathbf{B}\mathbf{r}(i)y^{*}(i) \big)^{H}\mathbf{B}\mathbf{r}(i)|=\sqrt{1+\gamma(i)},
 \end{split}
 \end{equation}
  which results in
the following
\begin{equation}
\mu(i)=\Big(1-\frac{\sqrt{1+\gamma(i)}}{|y(i)|}   \Big) \frac{1}{(\mathbf{r}^{H}(i)\mathbf{B}^{H}|y(i)|^{2}-\mathbf{r}^{H}(i)\mathbf{B}^{H})\mathbf{B}\mathbf{r}(i)}.\label{eq:mu1}
\end{equation}
For  case 2): $|y(i)|=| (v\mathbf{a}(\theta_{0})- \mathbf{B}^{H}\mathbf{w}(i) )^{H}\mathbf{r}(i)| < \sqrt{1-\gamma(i)}$,
$\mathbf{w}(i)$ is closer to the hyperplanes defined by $|(v\mathbf{a}(\theta_{0})- \mathbf{B}^{H}\mathbf{w} )^{H}\mathbf{r}(i)|=\sqrt{1-\gamma(i)}$.
By following the same approach we have
\begin{equation}
\mu(i)=\Big(1-\frac{\sqrt{1-\gamma(i)}}{|y(i)|}   \Big) \frac{1}{(\mathbf{r}^{H}(i)\mathbf{B}^{H}|y(i)|^{2}-\mathbf{r}^{H}(i)\mathbf{B}^{H})\mathbf{B}\mathbf{r}(i)}.\label{eq:mu2}
\end{equation}
Therefore, the expressions for $\mu(i)$ can be summarized in (\ref{eq:smcmgsc2}).

\end{document}